# History of Onychophorology, 1826-2020

# Historia de la Onicoforología, 1826-2020

**Julián Monge-Nájera,** julianmonge@gmail.com, Orcid ID: 0000-0001-7764-2966

Laboratorio de Ecología Urbana, Universidad Estatal a Distancia,  2050 San José, Costa Rica

## Abstract

Velvet worms, or onychophorans, include placental species and, as a phylum, have survived all mass extinctions since the Cambrian. They capture prey with an extraordinary adhesive net that appears in an instant.  The first naturalist to formally mention them was Lansdown Guilding (1797-1831), a British priest from the Caribbean island of Saint Vincent. His life is as little known as the history of the field he initiated, onychophorology. This is the first general history of onychophorology, and I have divided it into half century periods. The beginning, 1826-1879, was defined by former students of great names in the history of biology, like Cuvier and von Baer. This generation included  Milne-Edwars and Blanchard, and the greatest advances came from France, with smaller but still important contributions from England and Germany. In the 1880-1929 period, work concentrated in anatomy, behavior, biogeography and ecology, but of course the most important work was Bouvier's mammoth monograph. The next half century, 1930-1979, was important for the discovery of Cambrian species; Vachon's explanation of how ancient distribution defined the existence of two families; Pioneer DNA and electron microscopy from Brazil; and primitive attempts at systematics using embryology or isolated anatomical characteristics. Finally, the 1980-2020 period, with research centered in Australia, Brazil, Costa Rica and Germany, is marked by an evolutionary approach to everything, from body and behavior to distribution; for the solution of the old problem of how they form their adhesive net and how the glue works; the  reconstruction of Cambrian onychophoran communities, the first experimental taphonomy; the first country-wide map of conservation status (from Costa Rica); the first model of why they survive in cities; the discovery of new phenomena like food hiding, parental feeding investment and ontogenetic diet shift; and for the birth of a new research branch, Onychophoran Ethnobiology, founded in 2015. While a few names appear often in the literature, most knowledge was produced by a mass of researchers who entered the field only briefly.

## Resumen

Los gusanos de terciopelo, u onicóforos, incluyen especies con placenta y como grupo han sobrevivido a todas las extinciones masivas desde el Cámbrico. Capturan sus presas con una extraordinaria red adhesiva. El primer naturalista que los describió formalmente fue Lansdown Guilding (1797-1831), un sacerdote británico de la isla caribeña de San Vicente. Este documento es la primera historia general de la onicoforología. El comienzo, 1826-1879, fue definido por estudiantes de grandes nombres en la historia de la biología, como Cuvier y von Baer. Esta generación incluyó a Milne-Edwards y Blanchard, y los mayores avances vinieron de Francia. En el período 1880-1929, el trabajo se concentró en anatomía, comportamiento, biogeografía y ecología, pero el trabajo más importante fue la enorme monografía de Bouvier. El siguiente periodo, 1930-1979, fue importante por el descubrimiento de fósiles cámbricos; la explicación de Vachon de cómo la distribución antigua definió la existencia de dos familias; estudios pioneros en Brasil con ADN y microscopía electrónica; e intentos primitivos de sistemática utilizando embriología o características anatómicas aisladas. Finalmente, el período 1980-2020, con investigaciones centradas en Australia, Brasil, Costa Rica y Alemania, está marcado por un enfoque evolutivo de todos los campos, desde el cuerpo y el comportamiento, hasta la distribución geográfica; la solución del antiguo misterio de cómo forman su red adhesiva y cómo funciona el pegamento; la reconstrucción de las comunidades onicóforas del Cámbrico, la primera tafonomía experimental; el primer mapa del estado de conservación en todo el país (de Costa Rica); el primer modelo de porqué sobreviven en las ciudades; el descubrimiento de nuevos fenómenos como esconder alimentos, la inversión en alimentación parental y el cambio ontogenético de dieta; así como el nacimiento de una nueva rama de investigación, la etnobiología de onicóforos, fundada en 2015. Si bien algunos nombres aparecen a menudo en la literatura, la mayoría del conocimiento fue producido por una masa de investigadores que ingresaron al campo solo brevemente.

**Palabras clave:** historia de la ciencia, estudio de invertebrados, patrones de investigación, estudio de gusanos de terciopelo.

## INTRODUCTION

Velvet worms, or onychophorans, include placental species and, as a phylum, have survived all mass extinctions since the Cambrian. They capture prey with an extraordinary adhesive net that appears in an instant. There is not a single general history of this branch of science called onychophorology; the word does not even appear in dictionaries at the time I write this, but it has been used for decades by the Centre International de Myriapodologie in Paris. A definition is in order, so here is mine: ***Onychophorology* is a field of biology that studies all members of the phylum Onychophora and all subjects related to them in all fields of research**. These animals are extraordinary from the point of view of the evolution of life on Earth, but only interest a minuscule fraction of the scientific community. If all the onychophorologists active in the year 2020 were inside a bus, there would be many empty sits.

Here I summarize what are, in my opinion, the most significant contributions in the history of onychophorology. It is based on the *General Bibliography of Onychophora, 1826-2000*, available online here: https://zenodo.org/record/3698134#.XmEuBBP0nOQ

**A new phylum is discovered among humble plants: 1826 to 1879**

Other people probably saw onychophorans before Guilding (Costa Rican farmers refer to them as "slugs with legs"), but he was the first to describe them in a scientific paper (Guilding, 1826). It is hard to imagine the world in which he lived, where slavery was legal, Beethoven was still alive and no one knew that bacteria existed and caused disease. The Reverend Lansdown Guilding (1797-1831) was a British naturalist from the Caribbean island of Saint Vincent. He was mainly a botanist, a brilliant young man who corresponded with Darwin and Hooker; Hooker described him as "an arrogant, demanding, ambitious, and often conceited individual, all too ready to ask for unusual favors" (Howard & Howard, 1985). We know little else about Guilding, other than his first wife died "in childbed" leaving five children behind and that he died of unknown causes in 1831 while on vacation in another island (Howard & Howard, 1985).

Of the first and only onychophoran he ever saw, he wrote "it inhabits primary forests in Saint Vincent, often walks backward. If pressed, it releases viscous liquid from the mouth. Among the plants that I collected at the foot of mount "Bon Homme", I, astonished, discovered by chance the only specimen" (Monge-Nájera, 2019a).

Seven years after the publication, and two after Guilding's death, two French zoologists, Jean Victor Victoire Audouin (1797-1841) and Henri Milne-Edwards (1800-1885, a student of Georges Cuvier) moved onychophorans froms mollusks to annelids (Audouin & Milne-Edwards, 1833). Soon afterwards, the scientific world received the extraordinary news that the animal was also found half a world away, in South Africa (Gervais, 1836). Some pioneers in the field appear in Figure 1.

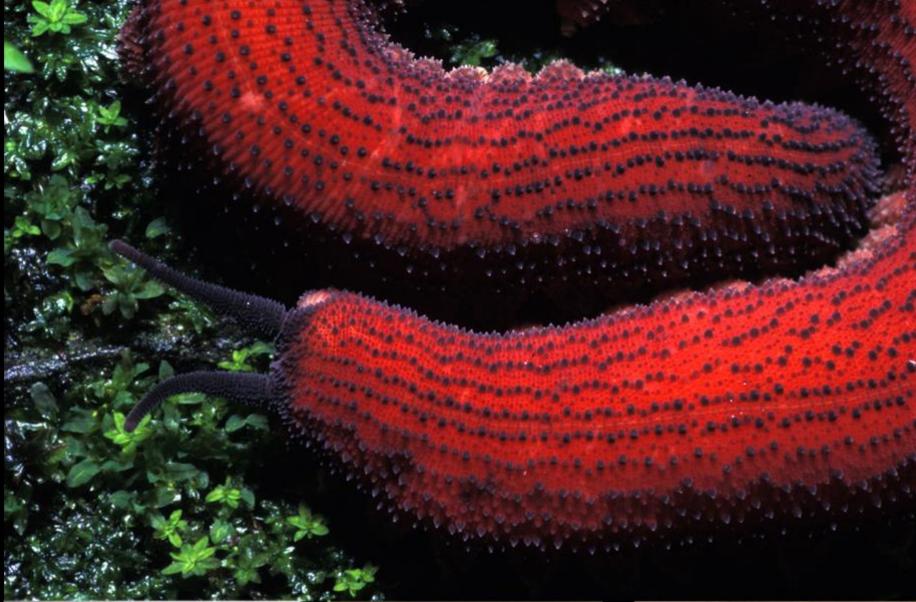

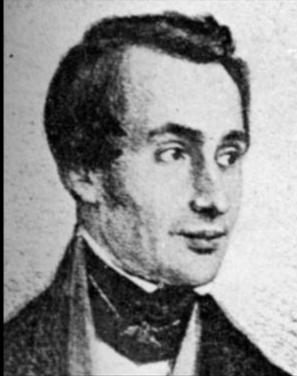 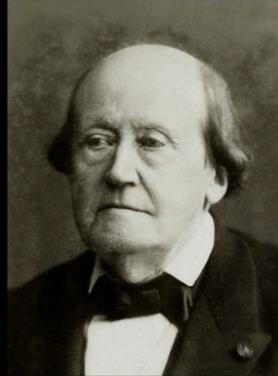 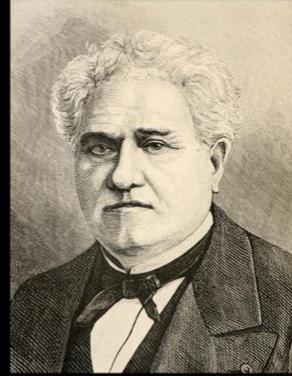

AUDOIN     MILNE     GERVAIS

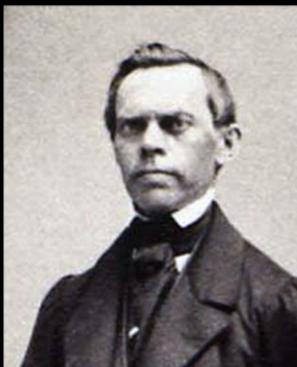 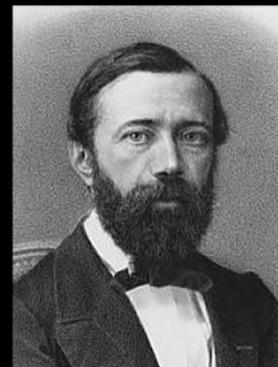 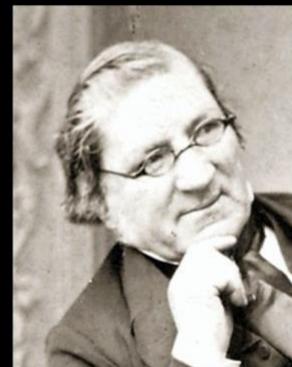

GRUBE     BLANCHARD     GAY

Fig. 1. A giant onychophoran from Costa Rica, and six pioneers of Onychophorology.  Source:
Alejandro Solórzano (worm) and, for the portraits, https://commons.wikimedia.org/

During the 1840s the animals were studied by classical French luminaries Jean de Quatrefages and Émile Blanchard (Quatrefages, 1848; Blanchard, 1847). Blanchard also wrote the onychophoran chapter for the monumental book series *Historia Física y Política de Chile* edited by Claude Gay (1800-1873) a French naturalist who pioneered natural history in Chile (Blanchard, 1849).

The raising of onychophorans to their own phylum was done in 1853 by Adolf Eduard Grube (1812-1880), then a lecturer in zoology in Dorpat, Germany. Grube, a student of the famous von Baer, father of embryology, was himself a recognized authority on invertebrates and specialized in Mediterranean polychaete worms (Grube, 1853).

The 1860s were poor in production, characterized by short notes, but for the first time in Europe (because it had been done earlier in Chile) we see the inclusion of the new animals in a general zoology, in this case, the *Leipzig Handbuch der Zoologie* (Carus, 1863).

The 1870s had a marked increase in productivity, mostly about anatomy, but also with a pioneering study about the embryology of *Peripatopsis capensis* from South Africa (Gegenbaur, 1874) and an early attempt on the evolutionary history of the group and its possible relationship with the origin of insects (Wood-Mason, 1879).

**They are all over the world! Bouvier enters the scene: 1880 to 1929**

The already notable growth in publications from the previous decade was followed by an even more spectacular increase in the 1880s, with papers on anatomy, embryology, ecology, geographic distribution and behavior. It was also the time of the oldest thesis on these animals that I could find: a study on the anatomy and histology of "peripatus" from the University of Breslau, in what is now Wrocław, Poland (Gaffron, 1883)

The embryology papers were about species from South America (Sclater, 1888), South Africa (Balfour, 1883; Sedgwick, 1884) and New Zealand (Sheldon, 1887). Others dealt with compared anatomy of the brain (Saint-Remy, 1889) and the farynx (Nicolas, 1889), the origin of metamerism (Sedgwick, 1884), and the first study focused on how the number of legs varies, this one from South Africa (Peters, 1880).

Ecological papers described the habitat of species from New South Wales (Bell, 1887) and New Zealand (Kirk, 1883), and reverend Adam Sedgwick, Darwin's geology teacher, published the first monograph of species and their geographic distribution (Sedgwick, 1908a). Smaller papers expanded the known distribution of the phylum in Asia and Oceania (e.g. Horst, 1886).

The Onychophora report from the *H.M.S.Challenger* was also published in this decade (Anonymous, 1885) as well as the first study about the animal's movements (Haase, 1889).

The decade of 1890 was marked by many natural history notes, as more specimens were found around the globe; there was less embryology and more ecology.

A study compared ovum development in South Africa and New Zealand (Sheldon, 1890) and Prenant (1890) described the seminal vesicles, which would prove important in understanding evolutionary pressures upon these animals (Monge-Nájera, 2019f), just like hypodermic impregnation, which was first described by Whitman (1891). Curiously, this decade also produced a pioneering work on the presence of corspuscles in the adhesive that the animal uses to hunt (Dendy, 1889); the adhesive would remain mostly forgotten as a study subject for over a century, until it became a leading-edge subject in the 21st century (Concha et al., 2015).

The first studies on onychophoran eggs and hatching (a still poorly known subject) were also written by Dendy (1889) and discussed by Fletcher (1891) in these late years of the 19th century.

Of particular interest is the fact that even at this early period, Caribbean onychophorans were so rarely seen that a note of the *rediscovery* of one was published by Grabham and Cockerell (1892), and that they were included in the reports of "noteworthy findings" by a natural history club that operated in the island of Trinidad (Anonymous, 1895).

Misidentification of species, still a problem in 2020, was mentioned as a problem over a century ago by Fletcher (1895) regarding Australian species. Other papers followed and dealt with the evolution of the onychophoran body (Goodrich, 1897) and their relationships with other invertebrates (Boas, 1898; Packard, 1898).

An unjustly forgotten author, Italian zoologist and alpinist Lorenzo Camerano (1856-1917), published several papers on the species of Panama and the Andes (e.g. Camerano, 1896), and around this time we also find the first papers by French zoologist Bouvier (Figure 2), the founding father of modern onychophorology, who dealt with their origin, evolution, variation and biogeography (e.g. Bouvier, 1902).

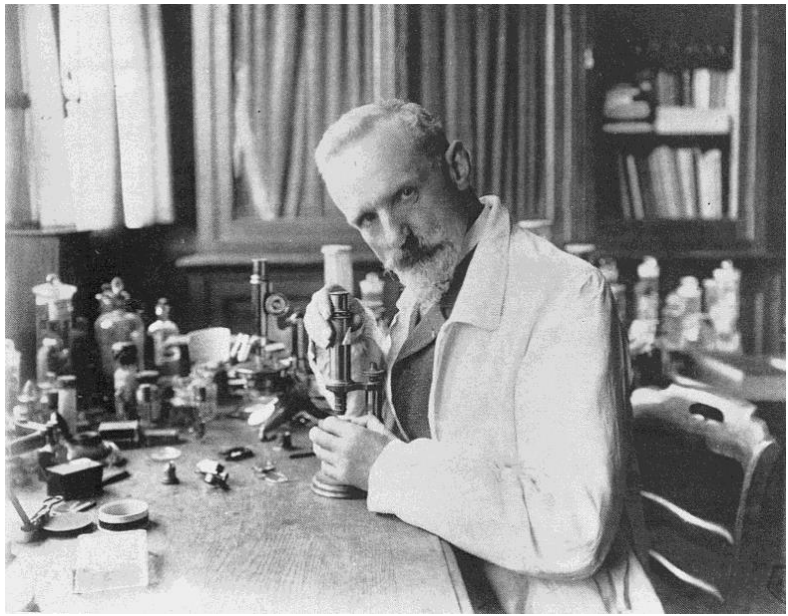

Fig. 2. Eugène Louis Bouvier.  Source:
https://commons.wikimedia.org/wiki/File:Louis_Eug%C3%A8ne_Bouvier.jpg

The 1900s started well, with the first paper on spermatogenesis (Montgomery, 1900) and also the first one on population density (Duerden, 1901). But undoubtedly what marks this period is the appearance of a large mass of literature by Bouvier, including evolution (Bouvier, 1902), and his monumental monograph (Bouvier, 1905, 1907), where he summarized all that was known about the animals, with emphasis on formal species descriptions. His species descriptions may be insufficient according to the standards of the 21st century, but are extraordinary if one considers the equipment and resources available to him at the time, and the fact that Eugène Louis Bouvier (1856-1944) was a busy man working mostly on crabs at the time. The son of a watchmaker, Bouvier moved up the social ladder through hard work, first as a teacher in a primary school, then by teaching about mosses and lichens in a Pharmacy School (Anonymous, 2012). It was only in 1895 that he got the chair of entomology at the National Museum of Natural History in Paris; the chair was previously occupied by Émile Blanchard (who had written the onychophoran section for the Chilean monograph mentioned earlier; Anonymous, 2012).

At the museum, Bouvier established what today would be called a "citizen science" program to enlarge the collections, and he also wrote a textbook of natural history for colleges that came out at the same time as his onychophoran monograph. He was then with the Prince of Monaco expedition, studying deep water crustaceans from the Sargasso Sea (Anonymous, 2012).

Later in his life, Bouvier wrote popular books about insect behavior (perhaps under pressure, because he had written little in the field of insects, despite keeping the entomology chair), but in science he is best remembered for his studies of crabs and onychophorans, which appeared to have been his real love (Blanckaert & Hurel, 2017).

This is also the decade in which researchers studied how onychophoran bodies process waste (Bruntz, 1903) and did more comparative studies, considering the parapodia of onychophorans and millipedes (Lankester, 1904); at the same time Sedgwick made the first attempt to associate the systematic relationships of onychophorans with their distribution around the world (Sedgwick, 1908a, b). Perhaps the most extraordinary products from this time were the lessons by C.E. Porter, in which he taught the natural history of onychophorans to students of the Naval Officers School of Chile, lessons that were later collected in the *Chilean Journal of Natural History* (Porter, 1905) and that probably are not part of the curriculum in naval schools today.

Their known distribution in Asia and Oceania was expanded in the decade of 1910 (Horst, 1910; Annandale, 1912); a curious observation on the discharge of mitochondria from the spermatozoon was reported (Montgomery, 1912) and Clark (1915) analyzed their world distribution, while the natural history and bibliography of the Chilean species were reviewed by two authors (Johow, 1911; Porter, 1917). An interesting finding from the time was that they could also be found in the forest canopy, according to Costa Rican microbiologist Clodomiro Picado (Picado, 1911).

The 1920s produced the first report of an onychophoran birth in captivity (for specimens kept in England: Dakin, & Fordham, 1926); and the first study of their diet, from Chile (Janvier, 1928). Bouvier published an "answer to Claude-Joseph" about Chilean species (Bouvier, 1928), but most of the work from this period focused on the head parts of onychophorans: the eye (Dakin, 1921); infra-cerebral organs (Dakin, 1922), ventral brain organs (Duboscq, 1920) and the evolution of the head in relationship with other invertebrates (Crampton, 1928). This period even has a rare Soviet contribution, a general morphology of the brain, read during the Second Congress of Zoologists, Anatomists, and Histologists of the USSR (Fedorov, 1927).

**A new, electronic look, and the role of DNA: 1930 to 1979**

The 1930s had the first papers on Cambrian onychophorans (Hutchinson, 1930; Walcott, 1931), as well as a study of the local variation of a species, a problem that troubled Bouvier and still troubles us (Brues, 1935). A rare report on the parasites of onychophorans (Vincent, 1936) has not, unfortunately, been followed by much work afterwards, leaving this as an almost virgin territory for exploration.

This period is marked by the first papers by authors that would become important in next decade, such as Marcus in Brazil and Manton in England (Marcus, 1937; Manton, 1937), as well as by Snodgrass' classic work on the compared evolution of onychophoran and arthropod bodies (Snodgrass, 1938).

The decade of 1940 is important because it has the first studies about a region that would later become a center of onychophoran research, Central America; the articles mostly resulting from strong research activity in the Panama canal area during, and after, World War II (Brues, 1941; Dunn, 1943; Clark & Zetek, 1946; Hilton, 1946; Arnett, 1947).

The 1950s was characterized by articles in popular magazines from London (Hill, 1950), South Africa (Lawrence, 1950), New Zealand (Wenzel, 1950), Germany (Zilch, 1955), New York (Milne & Milne, 1954; Alexander, 1958), Philadelphia (Bellomy, 1955), Malaysia (Hendrickson, 1957) and Belgium (Darteville, 1958).

Significant studies from the period considered how onychophorans produce leukocytes (Arvy, 1954), "muscle pharmacology" and possible uses of these animals in pharmacology (Ewer & van der Berg, 1954; Trindade, 1958); crural gland microanatomy (Gabe, 1956); brain secretions (Sanchez, 1958; Mendes & Sawaya, 1958); oxygen consumption in relation to size, temperature and oxygen tension (Mendes & Sawaya, 1958) and the formation of sperm cells (Tuzet & Manier, 1958; Gatenby, 1959). Perhaps the most innovative study from this decade was the first biogeographic analysis that explained the division into two families, as the results of historical separation at the time of Gondwana and Laurasia (Vachon, 1954).

Pioneering work in the 1960s originated mainly in Brazil, including the first study of DNA (Simoes, Marques da Silva, & Schreiber, 1964), the first observation with the electron microscope (Lavallard, 1965), and detailed studies on moulting (Campiglia, 1969). Reassessments of the fossils *Xenusion* (onychophoran or coelenterate?) and *Aysheaia* are also from this period (Tarlo, 1967; Hutchinson, 1969).

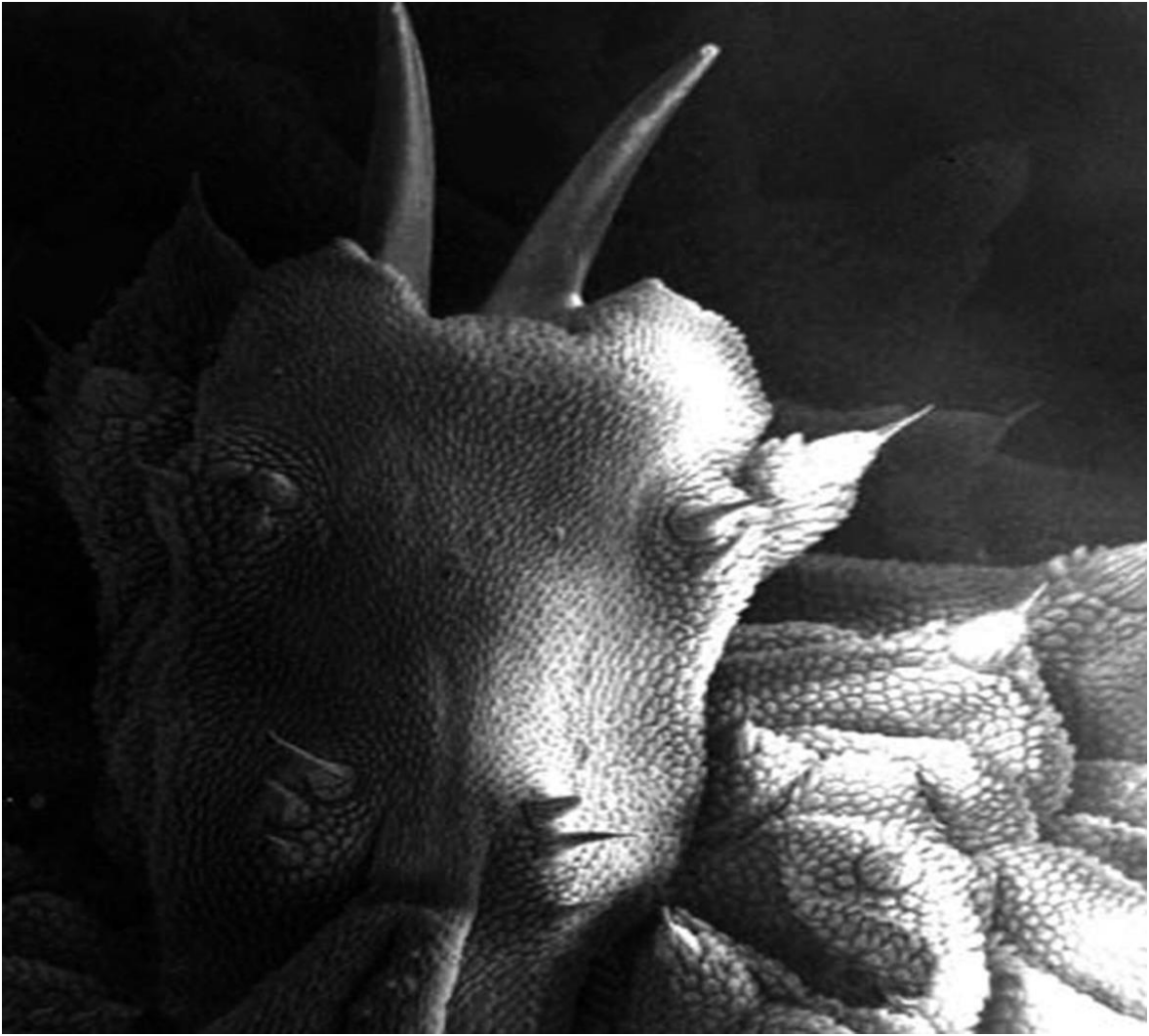

Fig. 3. The electron microscope presents the onychophoran surface more clearly than the light microscope and greatly facilitates taxonomic work. This is an example from the giant species that lives in the Caribbean of Costa Rica. Source Morera-Brenes, & Monge-Nájera (2010).

The 1970s were characterized by an extraordinary increase in anatomical and physiological work: the anatomy of the body wall (Birket-Smith, 1974), giant fibres in the ventral nerve cord (Schürmann & Sandeman, 1976), sarcoplasmic reticulum of smooth muscle (Heffron, Hepburn & Zwi, 1976), the sensilla (Storch & Ruhberg, 1977), the salivary glands (Ruhberg, 1979), and synaptic zones of nephrids (Storch, Alberti, Lavallard, & Campiglia, 1979).

Physiology studies analyzed cuticular chemistry and hardening (Krishnan, 1970); skeletal collagen (Hepburn & Heffron, 1976); enzymic activities of the smooth body-wall muscle (Heffron, Hepburn, & Zwi, 1977); the presence of monoamines in the nervous system (Gardner, Robson, & Stanford, 1978) and neuromuscular transmission (Hoyle & del Castillo, 1979).

Attempts were also made to disentangle systematics by comparing isolated aspects, such as locomotion (Manton, 1972), cuticle (Hackman & Goldberg, 1975), compared anatomy (de la Fuente, 1975) haemolymph (Gowri & Sundara Rajulu, 1976), and the Golgi complex (Locke & Huie, 1977). These attempts had little chance of succeeding because the resulting phylogenetic trees produced by any particular character were incompatible with the trees produced by other characters.

This is also the time of the first general study about the habitat (Lavallard, Campiglia, Parisi Alvares, & Valle, 1975). Lawrence (1977) summarized research in South Africa, while Peck (1975) reviewed the species of the American continent and Delle Cave and Simonetta (1975) added new information on the morphology and taxonomic position of the fossil *Aysheaia*.

**An exciting period of unveiled secrets and conservation worries: 1980 to 2020**

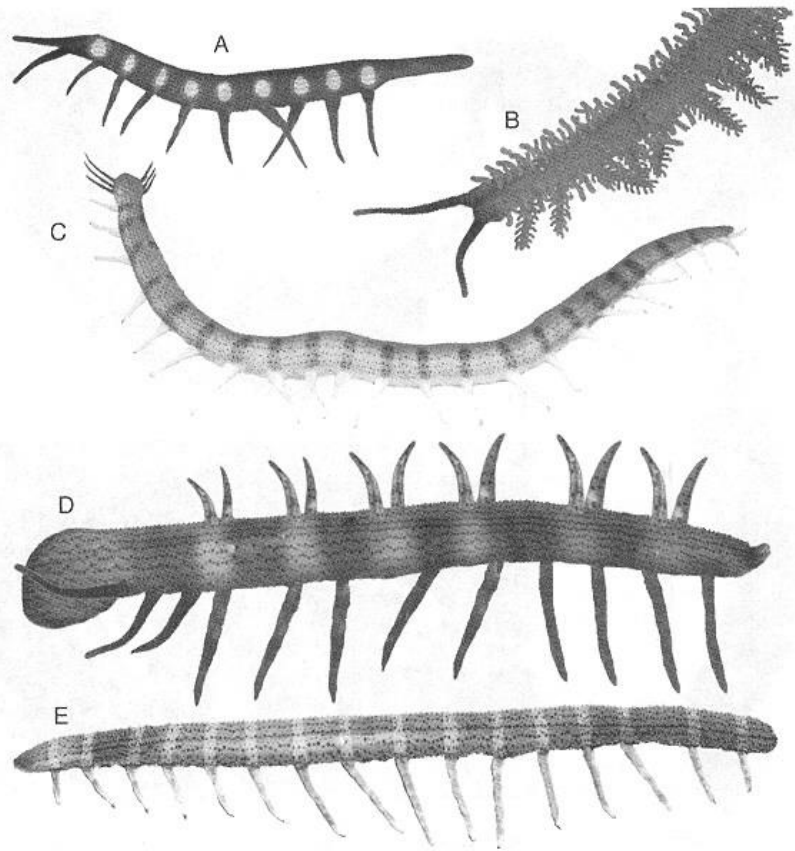

Fig. 4. Digital reconstructions of fossil onychophorans (Monge-Nájera & Xianguang, 2002). A: *Microdictyon sinicum*, B: *Onychodictyon ferox*, C: *Cardiodictyon catenulum*, D: *Hallucigenia fortis*, E: *Luoishania lingicruris*

Much study was done in the 1980s on the anatomy of onychophorans: musculature and innervation (Hoyle and Williams, 1980); morphometry of the tracheal system (Bicudo and Campiglia, 1985); the application of scanning electron microscopy to systematics (Read, 1988); and the curious cephalic pits and palps of some Australian species (Ruhberg, Tait, Briscoe, & Storch, 1988).

This period is also marked by the first strongly evolutionary focus on their reproduction (Morera, Monge-Nájera, & Saenz, 1988; Havel, Wilson, & Hebert, 1989), and seems to also be the first time that onychophorans appear in the *IUCN Invertebrate Red Data Book* (Wells, Pyle, & Collins, 1983). New possible onychophoran fossils were reported from Illinois (Thompson & Jones, 1980) and France (Rolfe, Schram, Pacaud, Sotty, & Secretan, 1982), but perhaps the most important publication from this decade is the second monograph on the family Peripatopsidae (Ruhberg, 1985).

The 1990s was characterized by research in genetics and systematics. RNA and DNA sequences were combined with morphological data to assess systematics, using antennal circulatory organs (Pass, 1991; de Haro, 1998), and a paper recommended rejection of the "Uniramia" hypothesis, in which Sidney Manton joined Hexapoda, Myriapoda and Onychophora as a single monophyletic group (Wägele, 1993).

Overall, researchers in several regions of the world discovered the large genetic variation hidden in what seemed to be a relatively simple phylum (Grenier, Garber, Warren, Whitington, & Carroll, 1997; Curach & Sunnucks, 1999; Hebert et al., 1991; Morera-Brenes, Herrera, Mora, & Leon,1992; Ballard et al., 1992; Briscoe & Tait, 1995; Gleeson, 1996).

The second area of this decade, by productivity, was ecology and biogeography. Findings from Singapore showed that recent introductions could complicate the reconstruction of onychophoran historical biogeography (van der Lande, 1991), and pioneering computerized techniques were used to understand how climate and paleo-vegetation define their current world distribution (Monge-Nájera, 1994a), as well as how climate affects body characteristics (Monge-Nájera, 1994b)

This decade was also marked by the first detailed study of the geographic variation of habitats (Monge-Nájera & Alfaro, 1995), and a model of onychophoran biogeographic history from the Jurassic through the Pliocene (Monge-Nájera, 1996). Other salient studies were a comparative analysis of evolutionary trends in onychophorans and scorpions (Monge-Nájera & Lourenco, 1995), and the first in-depth studies of the onychophoran adhesive secretion, with electrophoresis (Mora, Herrera, & León, 1996); chemical characterization (Benkendorff, Beardmore, Gooley, Packer, & Tait, 1999) and a comparison with arthropod silk secretions (Craig, 1997).

In the field of conservation, the finding of a single large aggregation in New Zealand (Harris, 1991) and a unconfirmed estimate that the population of a single species reached millions, led to the unjustified generalization that these animals "are not rare" (Mesibov, 1998), an error that can lead to dangerous implications for conservation (Monge Nájera, 2019d).

In ethology, this period is marked by the first experimental study of general behavior (Monge-Nájera, Barrientos, & Aguilar, 1993) and of the pheromonal function of crural glands (Eliott, Tait, & Briscoe, 1993).

In fossils, the most important findings were the reinterpretation of *Hallucigenia*, now thought to be an onychophoran that was originally interpreted upside-down by Morris (1977) and corrected by Hou and Bergström (1991); and the opposite case, a Palaeozoic "onychophoran" reinterpreted as a stalked echnoderm (Rhebergen & Donovan, 1994). But perhaps the most unfortunate error of the time was G. Poinar's creation of new "onychophoran families" with specimens that lacked the body parts needed to define families (Poinar, 1996, 2000).

This decade was also marked by the first "modern synthesis" that proposed evolutionary explanations for the origin of all known onychophoran characteristics and summarized their history since the Cambrian, including anatomy, physiology, behavior, distribution, reproduction and systematics (Monge-Nájera, 1995).

The decade of 2000 had an eclectic production. Papers dealt with how population structure and genetic constitution are related (Laat, 2006; Santana, Almeida, Alves, & Vasconcellos, 2008); the rediscovery, after more than a century, of *Oroperipatus eisenii* in Mexico (Cupul-Magaña & Navarrete-Heredia, 2008); and a revival of the ancient debate about the origin of head parts (e.g. Eriksson & Budd, 2003; Mayer & Whitington, 2009).

In paleoecology, a quantitative study of an exquisitely preserved Cambrian community from China found extraordinary similarities with an extant community in Costa Rica (Monge-Nájera & Hou, 2000), and experimental taphonomy identified which fossil onychophoran "structures" are real and which can be just artifacts (Monge-Nájera & Xianguang, 2002).

Other outstanding reports included the use the head to insert the spermatophore (Tait & Norman, 2001) and that female-dominated hierarchies (Reinhard & Rowell, 2005).

This decade was also marked by the first study of the inflammatory process in onychophorans (Silva, Coelho, & Nogueira, 2000) and the identification of immune inducible genes (Altincicek & Vilcinskas, 2008), but perhaps the most innovative work was the first study of the physics of the adhesive secretion (Jerez-Jaimes & Bernal-Pérez, 2009).

The most recent decade, the 2010s, has been marked by the discovery of the largest species ever and by the first study of the mechanism by which onychophorans produce their adhesive "hunting nets" (Morera-Brenes & Monge-Nájera, 2010). The adhesive skin exudate of some frogs was found to be similar to that of onychophorans (Graham, Glattauer, Li, Tyler, & Ramshaw, 2013), and the adhesive net mechanism was finally cracked and found to be produced by passive hydrodynamic instability (Concha et al., 2015).

Other studies from this time dealt with the nanostructures of the solidified adhesive secretion (Corrales et al., 2017), the assembling of fibers by electrostatic interactions in phosphoproteins (Baer, Hänsch, Mayer, Harrington, & Schmidt, 2018), and a mimic of onychophoran skin, which is slippery to the adhesive, produced with microporous porphyrin networks (Ryu et al., 2018).

In the field of genetics, the genome size and chromosome numbers were reviewed (Jeffery, Oliveira, Gregory, Rowell, & Mayer, 2012), and new morphological characters were added to the arsenal used in taxonomy (Oliveira, Read, & Mayer, 2012). Other studies dealt with genes related to head and eye development (Eriksson, Samadi, & Schmid, 2013); endoderm marker-genes during gastrulation and gut-development (Janssen & Budd, 2017); the conserved and derived cell death in embryonic development (Treffkorn & Mayer, 2017); and the report that fluorescence *in situ* hybridization of telomers indicate chromosome fusions (Dutra, Cordeiro, & Araujo, 2018).

In this decade, onychophorans were found in unexpected places, like Vietnam (Bái & Anh, 2012), lava tubes (Espinasa et al., 2015) and urban vegetation (Barrett, 2013), and the first model to explain and predict their survival in highly disturbed habitats, based on their size and habits, was published (Monge-Nájera, 2018). The only tropical African species was collected again after more than a century (Costa & Giribet, 2016), and the period was also marked by advances in the study of onychophoran genetics and conservation in Brazil (Lacorte, De Sena Oliveira, and Da Fonseca, 2011; Costa, 2016; Cunha et al., 2017; Costa et al., 2018).

Costa Rica became the first country to fully map the distribution of its onychophorans and to indicate where they were preserved and the strength of conservation measures for each species (Morera, Monge-Nájera, & Mora, 2018). The first field monitoring study covering *several years* found that the relationship between onychophoran hunting activity, humidity and light, was not as expected (Barquero-González, Morera-Brenes, & Monge-Nájera, 2018). Food hiding, parental feeding investment and ontogenetic diet shift were also reported for the first time for the whole phylum (Barquero-González, Vega-Hidalgo, & Monge-Nájera, 2019).

A new branch of onychophorology was also born in this decade, the Ethnobiology of Onychophorans, with the first study of onychophoran representations in folklore and art (Monge-Nájera & Morera-Brenes, 2015). The period ends with a series that proposes evolutionary explanations for several previously unexplained anatomical, physiological, behavioral and ecological characteristics, such as why do onychophoran spermatozoa swim for years, why do some Australian onychophorans have bizarre heads, why are there no onychophorans in Cuba and why ovoviviparity may be the ancestral form of reproduction in velvet worms (Monge-Nájera, 2019b, 2019c, 2019d, 2019e, 2019f).

In a period of climatic change and fear of mass extinctions, perhaps the most significant paper of the decade may be one that presents a way to rapidly and cheaply distinguish and name onychophoran species for their protection and long term conservation (Sosa-Bartuano, Monge-Nájera, & Morera-Brenes, 2018).

**CONCLUSION**

In these two centuries, Onychophorology has been built by contributions from many men and women. A few stayed long enough in the field to write numerous contributions (e.g. Barquero-González, Bouvier, Campiglia, Daniels, Mayer, Monge-Nájera, Morera-Brenes, Lavallard, Oliveira, Ruhberg, Sedgwick, Storch, Sunnucks, and Tait), but we ought the bulk of our knowledge to the mass of men and women who entered the field only briefly. There is a lesson here: for onychophorology to prosper, we need to attract as many researchers from other fields as possible, even if each produces only one paper, for it is in their brains that the future of onychophoroly will be born.

## ACKNOWLEDGEMENTS

This study was financed by the author. I thank Carolina Seas and Bernal Morera for their assistance with format, illustrations and information.

## AUTHOR CONTRIBUTION STATEMENT

The total contribution percentage for the conceptualization, preparation, and correction of this paper was J.M.N. 100 %.

## DATA AVAILABILITY STATEMENT

The data supporting the results of this study is fully and freely available here: https://zenodo.org/record/3698134#.XmEuBBP0nOQ.

## REFERENCES

[Anonymous]. (1885). Peripatus. In *Report on the Scientific Results of the Voyage of H.M.S. Challenger During the Years 1873-*76 (pp. 284-286). London: Longmans & Co.

[Anonymous]. (1895). Report of club meetings, 19 April 1895. *Journal of the Trinidad Field Naturalists' Club, 2*, 187-189.

[Anonymous]. (2012). *Eugène Louis Bouvier (1856-1944).* Paris: Institut Pasteur. Retrieved from https://webext.pasteur.fr/archives/buv0.html

Alexander, A. J. (1958). Peripatus: Fierce little giant. *Animal Kingdom, 61*, 122-125.

Altincicek, B., & Vilcinskas, A. (2008). Identification of immune inducible genes from the velvet worm Epiperipatus biolleyi (Onychophora). *Developmental and Comparative Immunology, 32*(12), 1416-21.

Annandale, N. (1912). The occurrence of Peripatus on the North-East frontier of India. *Nature, 88*, 449.

Arnett, R. H. (1947). Epiperipatus braziliensis (Bouvier) on Barro Colorado Island, Canal Zone. *Entomological News, 58*, 59-60.

Arvy, L. (1954). Présentation de documents sur la leucopoïèse chez Peripatopsis capensis Grube. *Bulletin de la Société Zoologique de France, 79*, 13.

Audouin, M. & Milne-Edwards, H. (1833). Classification des Annélides, et description de celles qui habitent les cites de la France - sixième Famille: Péripatiens. *Annales des Sciences Naturelles, 30*, 411-414.

Baer, A., Hänsch, S., Mayer, G., Harrington, M. J., & Schmidt, S. (2018). Reversible Supramolecular Assembly of Velvet Worm Adhesive Fibers via Electrostatic Interactions of Charged Phosphoproteins. *Biomacromolecules, 19*(10), 4034-4043. DOI: 10.1021/acs.biomac.8b01017

Bai, T. T., & Anh, N. D. (2012). Discovery of Eoperipatus sp. (Peripatidae), the first representative of Onychophora in Vietnam. *TAP CHI SINH HOC, 32*(4), 36-39. DOI: 10.15625/0866-7160/v32n4.718

Balfour, F.M. (1883). The anatomy and development of Peripatus capensis. *Quarterly Journal of Microscopical Science, 23*, 213-259.

Ballard, J. W. O., Olsen, G. J., Faith, D. P., Odgers, W. A., Rowell, D. M. & Atkinson, P. W. (1992). Evidence from 12S ribosomal RNA sequences that onychophorans are modified arthropods. *Science, 258*, 1345-1348.

Barquero-González, J. P., Morera-Brenes, B., & Monge-Nájera, J. (2018). The relationship between humidity, light and the activity pattern of a velvet worm, Epiperipatus sp. (Onychophora: Peripatidae), from Bahía Drake, South Pacific of Costa Rica. *Brazilian Journal of Biology, 78*(3), 408-413. DOI: 10.1590/1519-6984.166495

Barquero-González, J. P., Vega-Hidalgo, A., & Monge-Nájera, J. (2019). Feeding behavior of Costa Rican velvet worms: food hiding, parental feeding investment and ontogenetic diet shift (Onychophora: Peripatidae). *UNED Research Journal, 11*(2), 85-88. DOI: 10.22458/urj.v11i2.2195

Barrett, D. (2013). *Multiple-scale resource selection of an undescribed urban invertebrate (Onychophora: Peripatopsidae) in Dunedin, New Zealand* (Master´s Thesis). New Zealand, University of Otago. Retrieved from https://ourarchive.otago.ac.nz/handle/10523/4044

Bell, F. J. (1887). Habitat of Peripatus leuckarti. *Annals and Magazine of Natural History [Series 5], 20*, 252.

Bellomy, M. D. (1955). Peripatus - between worm and insect. *Frontiers (Philadelphia, Pennsylvania, USA), 20*, 42-44.

Benkendorff, K., Beardmore, K., Gooley, A. A., Packer, N. H. & Tait, N. N. (1999). Characterisation of the slime gland secretion from the peripatus, Euperipatoides kanangrensis (Onychophora: Peripatopsidae). *Comparative Biochemistry and Physiology, Part B, 124*, 457-465.

Bicudo, J. E. P. W., & Campiglia, S. (1985). A morphometric study of the tracheal system of Peripatus acacioi Marcus and Marcus (Onychophora). *Respiration Physiology, 60*, 75-82.

Birket-Smith, S. J. R. (1974). The anatomy of the body wall of Onychophora. *Zoologische Jahrbücher, Abteilung für Anatomie und Ontogenie der Tiere, 93*, 123-154.

Blanchard, E. (1847). Recherches sur l'organisation des Vers. *Annales des Sciences Naturelles [3e Série], 8*, 119-149.

Blanchard, E. (1849). Malacopodes. In C. Gay (Ed.), *Historia Fisica y Politica de Chile (Vol. 23): Zoologia, Tomo Tercero* (pp. 57-60). Paris: privately published.

Blanckaert, C. & Hurel, A. (2017). *Louis-Eugène Bouvier (1856-1944), un amateur à la chaire d'entomologie?* Retrieved from https://objethistoire.hypotheses.org/953

Boas, I. E. V. (1898). Om Peripatus stilling i dyreriget. *Kongelige Danske Videnskabernes Selskabs Forhandlinger, 6*, 345-365.

Bouvier, E. L. (1902). Observations nouvelles sur l'évolution et l'origine des Péripates. *Comptes Rendus Hebdomadaires des Seances de l'Academie des Sciences, 134*, 55-58.

Bouvier, E. L. (1905). Monographie des Onychophores. *Annales des Sciences Naturelles, Zoologie et Biologie Animale [9e Série], 2*, 1-383.

Bouvier, E. L. (1907). Monographie des Onychophores. *Annales des Sciences Naturelles, Zoologie et Biologie Animale [9e Série], 5*, 61-318.

Bouvier, E. L. (1928). A propos des observations du Fr. Claude-Joseph sur un Péripate du Chili. *Annales des Sciences Naturelles, Zoologie et Biologie Animale [10e Série], 11*, 260.

Briscoe, D. A., & Tait, N. N. (1995). Allozyme evidence for extensive and ancient radiations in Australian Onychophora. *Zoological Journal of the Linnean Society, 114*, 91-102.

Brues, C. T. (1935). Varietal forms of Peripatus in Haiti. *Psyche, 42*, 58-62.

Brues, C. T. (1941). Peripatus (Macroperipatus) geayi in Panama. *Psyche, 48*, 111-112.

Bruntz, L. (1903). Excrétion et phagocytose chez les Onychophores. *Comptes Rendus Hebdomadaires des Seances de l'Academie des Sciences, 136*, 1148-1150.

Camerano, L. (1896). Onicofori raccolti nel Darien da Dott. E. Festa. *Bollettino del Museo Regionale di Scienze Naturali Torino, 11*, 1-2

Campiglia, S. S. (1969). *Tegumento, muda e ciclo de intermuda em Peripatus acacioi Marcus e Marcus (onicoforo)* (Dissertation thesis). Universidade de Sao Paulo, Brazil.

Carus, J. V. (1863). Onychophora. In J. V. Carus & A. Gerstaecker (Eds.), *Handbuch der Zoologie* (Vol. 2, pp. 446-447). Leipzig, Germany: Verlag von Wilhelm Engelmann.

Clark, A. H. & Zetek, J. (1946). The onychophores of Panama and the Canal Zone. *Proceedings of the United States National Museum, 96*, 205-213.

Clark, A. H. (1915). The present distribution of the Onychophora, a group of terrestrial invertebrates. *Smithsonian Miscellaneous Collections, 65*, 1-25.

Concha, A., Mellado, P., Morera-Brenes, B., Sampaio Costa, C., Mahadevan, L., & Monge-Nájera, J. (2015). Oscillation of the velvet worm slime jet by passive hydrodynamic instability. *Nature Communications, 6*, 6292.

Corrales-Urena, Y. R., Sanchez, A., Pereira, R., Rischka, K., Kowalik, T., & Vega-Baudrit, J. (2017). Extracellular micro and nanostructures forming the velvet worm solidified adhesive secretion. *Materials Research Express, 4*(12). DOI: 10.1088/2053-1591/aa9940

Costa, C. S. (2016). *Sistemática e análise filogenética de Epiperipatus Clark, 1913 baseada em dados moleculares e morfológicos (Onychophora: Peripatidae)* (Tese de Doutorado). Universidade de São Paulo, São Paulo. DOI: 10.11606/T.41.2016.tde-25082016-100135

Costa, C. S., Chagas-Junior, A., Pinto-da-Rocha, R., Costa, C. S., Chagas-Junior, A., & Pinto-da-Rocha, R. (2018). Redescription of Epiperipatus edwardsii, and descriptions of five new species of Epiperipatus from Brazil (Onychophora: Peripatidae). *Zoologia, 35*, 1-15. DOI: 10.3897/zoologia.35.e23366

Costa, C. S., & Giribet, G. (2016). Taxonomic Notes on Mesoperipatus tholloni (Onychophora: Peripatidae), an Elusive Velvet Worm from Gabon. *Breviora, 552*(1), 1-10. DOI: 10.3099/MCZ30.1

Craig, C. L. (1997). Evolution of arthropod silks. *Annual Review of Entomology, 42*, 231-267.

Crampton, G. C. (1928). The evolution of the head region in lower arthropods and its bearing upon the origin and relationships of the arthropodan groups. *Canadian Entomologist, 60*, 284-301.

Cunha, W. T. R., Santos, R. C. O., Araripe, J., Sampaio, I., Schneider, H., Rêgo, P. S., … Rêgo, P. S. (2017). Molecular analyses reveal the occurrence of three new sympatric lineages of velvet worms (Onychophora: Peripatidae) in the eastern Amazon basin. *Genetics and Molecular Biology, 40*(1), 147-152. DOI: 10.1590/1678-4685-gmb-2016-0037

Cupul-Magaña, F. G., & Navarrete-Heredia, J. (2008). Rediscovery and New Data for Oroperipatus eisenii (Wheeler, 1898) from Mexico (Onychophora: Peripatidae). *Entomological News, 119*(5), 545-549. DOI: 10.3157/0013-872x-119.5.545

Curach, N., & Sunnucks, P. (1999). Molecular anatomy of an onychophoran: Compartmentalized sperm storage and heterogeneous paternity. *Molecular Ecology, 8*, 1375-1385.

Dakin, W. J., & Fordham, M. G. C. (1926). Birth of Peripatus in England. *Nature, 117*, 858.

Dakin, W. J. (1921). The eye of Peripatus. Quarterly Journal of Microscopical. *Science, 65*, 163-172.

Dakin, W. J. (1922). The infra-cerebral organs of Peripatus. Quarterly *Journal of Microscopical Science, 66*, 409-417.

Darteville, E. (1958). Les Péripates. *Congo-Tervuren, 2*, 65-68.

de Haro, A. (1998). Origen y relaciones fitogenéticas entre Artrópodos, Onicóforos, Anélidos y Lofoforados, según datos moleculares y morfológicos. *Boletín de la Real Sociedad Española de Historia Natural Sección Biológica, 94*, 103-113.

de la Fuente, J. A. (1975). Esquema filogenético de la línea Onychophora-Myriapoda-Hexapoda. *Boletín de la Real Sociedad Española de Historia Natural Sección Biológica, 73*, 85-97.

de Quatrefages, A. (1848). Études sur les types inférieurs de l`embranchment des annelés. Mémoire sur la famille des Hermelliens (Hermellea Nob.). *Annales des Sciences Naturelles [3e Série], 10*, 5-58.

Delle Cave, L., & Simonetta, A. M. (1975). Notes on the morphology and taxonomic position of Aysheaia (Onychophora?) and of Skania (undetermined phylum). *Monitore Zoologico Italiano [Nuéva Sèrie], 9*, 67-81.

Dendy, A. (1889). Peripatus in Victoria. *Victorian Naturalist. 5,* 134-135.

Duboscq, O. (1920). Notes sur Opisthopatus cinctipes Purc. I. Sur les poils des papilles primaires et leur développement - II. Les organes ventraux du cerveau. *Archives de Zoologie Experimentale et Generale [Notes et Revue], 59*, 21-27.

Duerden, J. E. (1901). Abundance of Peripatus in Jamaica. *Nature, 63*, 440-441.

Dunn, E. R. (1943). Zoological results of the Azuero Peninsula Panama Expedition of 1940. Part I - A new species of Peripatus. *Notulae Naturae, 123*, 1-5.

Dutra, D. D., Cordeiro, L. M., & Araujo, D. (2018). Maior número cromossômico em Onychophora e indício de fusão cromossômica detectada por FISH telomérica. *Semina: Ciências Biológicas e Da Saúde, 38*(1supl), 199. DOI: 10.5433/1679-0367.2017V38N1SUPLP199

Eliott, S., Tait, N. N., & Briscoe, D. A. (1993). A pheromonal function for the crural glands of the onychophoran Cephalofovea tomahmontis (Onychophora: Peripatopsidae). *Journal of Zoology, 231*, 1-9.

Eriksson, B. J., & Budd, G. E. (2003). *The cephalic nerves of the Onychophora and their bearing on our understanding of head segmentation and stem-group evolution of Arthorpoda.* Retrieved from http://www.diva-portal.org/smash/record.jsf?pid=diva2%3A68261&dswid=-8030

Eriksson, B. J., Samadi, L., & Schmid, A. (2013). The expression pattern of the genes engrailed, pax6, otd and six3 with special respect to head and eye development in Euperipatoides kanangrensis Reid 1996 (Onychophora: Peripatopsidae). *Development Genes and Evolution, 223*(4), 237-246. DOI: 10.1007/s00427-013-0442-z

Espinasa, L., Garvey, R., Espinasa, J., Fratto, C. A., Taylor, S. J., Toulkeridis, T., & Addison, A. (2015). Cave dwelling Onychophora from a lava tube in the Galapagos. *Subterranean Biology, 15*(1), 1-10. DOI: 10.3897/subtbiol.15.8468

Ewer, D.W., & van der Berg, R. (1954). A note on the pharmacology of the dorsal musculature of Peripatopsis. *Journal of Experimental Biology, 31*, 497-500.

Fedorov, B. G. (1927). On the morphology of the brain of Peripatus. In N. Sewertzoff & B. S. Matveiev (Eds.), *Proceedings of the Second Congress of Zoologists, Anatomists, and Histologists of USSR: Moscow 4-10 May 1925* (pp. 92-94). Moscow: "Glavnauka".

Fletcher, J. J. (1891). Note on the supposed oviparity of P. leuckartii. *Proceedings of the Linnean Society of New South Wales, 2nd Series, 6*, 577.

Fletcher, J. J. (1895). On the specific identity of the Australian Peripatus, usually supposed to be P. leuckarti, Saenger. *Proceedings of the Linnean Society of New South Wales, 2nd Series, 10*, 172-194.

Gabe, M. (1956). Particularités histologiques des glandes crurales des Onychophores. *Bulletin de la Société Zoologique de France, 81*, 170.

Gaffron, E. (1883). *Anatomie und Histologie von Peripatus* (Dissertation thesis). Poland, University of Breslau.

Gardner, C. R., Robson, E. A., & Stanford, C. (1978). The presence of monoamines in the nervous system of Peripatopsis (Onychophora). *Experientia, 34*, 1577-1578.

Gatenby, J. B. (1959). A note on the spermiogenesis of Peripatoides novae-zealandiae. *Transactions of the Royal Society of New Zealand, 87*, 51-53.

Gegenbaur, C. (1874). Dritter Abschnitt. Würmer. In *Grundriss der Vergleichenden Anatomie Verlag von Wilhelm Engelmann, Leipzig* (pp. 122-204).

Gervais, P. (1836). Note déscriptive sur Peripatus brevis. *Bulletin de la Société Entomologique de France [1836], 15*.

Gleeson, D. (1996). Uncovering the radiation of New Zealand's Onychophora using molecular data. *Proceedings of the International Congress of Entomology, 19*, 61.

Goodrich, E. S. (1897). On the relation of the arthropod head to the annelid prostomium. *Journal of Cell Science, s2-40*, 247-268.

Gowri, N., & Sundara Rajulu, G. (1976). A comparative study of the organic components of the haemolymph of Eoperipatus weldoni (Onychophora) and a pill-millipede Arthrosphaera lutescens (Diplopoda). *Journal of Animal Morphology and Physiology, 23*, 60-75.

Grabham, M., & Cockerell, T. D. A. (1892). Peripatus re-discovered in Jamaica. *Nature, 46*, 514.

Graham, L. D., Glattauer, V., Li, D., Tyler, M. J., & Ramshaw, J. A. M. (2013). The adhesive skin exudate of Notaden bennetti frogs (Anura: Limnodynastidae) has similarities to the prey capture glue of Euperipatoides sp. velvet worms (Onychophora: Peripatopsidae). *Comparative Biochemistry and Physiology. Part B, Biochemistry & Molecular Biology, 165*(4), 250-9. DOI: 10.1016/j.cbpb.2013.04.008

Grenier, J. K., Garber, T. L., Warren, R., Whitington, P. M., & Carroll, S. (1997). Evolution of the entire arthropod Hox gene set predated the origin and radiation of the onychophoran/arthropod clade. *Current Biology, 7*, 547-553.

Grube, E. (1853). Über den Bau von Peripatus edwardsii. *Müller's Archives of Anatomy and Physiology [1853]*, 322-360.

Guilding, L. (1826). Mollusca Caribbaeana. *Zoological Journal, 2*, 437-449.

Haase, E. (1889). Über die Bewegungen von Peripatus. *Sitzungsberichte der Gesellschaft Naturforschender Freunde zu Berlin [1889]*, 148-151.

Hackman, R. H., & Goldberg, M. (1975). Peripatus: its affinities and its cuticle. *Science, 190*, 582-583.

Harris, A. C. (1991). A large aggregation of Peripatoides novaezealandiae (Hutton, 1876) (Onychophora: Peripatopsidae). *Journal of the Royal Society of New Zealand, 21*, 405-406.

Havel, J. E., Wilson, C. C., & Hebert, P. D. N. (1989). Parental investment and sex allocation in a viviparous onychophoran. *Oikos, 56*, 224-232.

Hebert, P. D. N., Billington, N., Finston, T. L., Boileau, M. G., Beaton, M. J., & Barrette, R. J. (1991). Genetic variation in the onychophoran Plicatoperipatus jamaicensis. *Heredity, 67*, 221-229.

Heffron, J. J. A., Hepburn, H. R., & Zwi, J. (1976). On the sarcoplasmic reticulum of onychophoran somatic smooth muscle. *Naturwissenschaften, 63*, 95.

Heffron, J. J. A., Hepburn, H. R. & Zwi, L. J. (1977). Enzymic activities of the smooth body-wall muscle of the onychophoran Peripatopsis moseleyi. *Biochemical Society Transactions, 5*, 1748-1750.

Hendrickson, J. (1957). Peripatus in Malaysia. *Malayan Nature Journal, 12*, 33-35.

Hepburn, H. R. & Heffron, J. J. A. (1976). On the skeletal collagen of an onychophoran, Peripatopsis mosleyi. *Cytobiologie, 12*, 481-486.

Hill, R. P. (1950). Peripatus: A missing link. *Discovery [London], 11*, 14-18.

Hilton, W. A. (1946). Remarks on the habitat of Peripatus on Barro Colorado Island, C.Z. *Journal of Entomology and Zoology [Claremont, California], 38*, 27.

Horst, R. (1886). On a specimen of Peripatus, Guild., from Sumatra. *Notes from the Leyden Museum, 8*, 37-41.

Horst, R. (1910). Paraperipatus lorentzi Horst, a new Peripatus from Dutch New Guinea. *Notes from the Leyden Museum, 32*, 217-218.

Hou, X. G. & Bergström, J. (1991). The arthropods of the Lower Cambrian Chengjiang fauna, with relationships and evolutionary significance. In A. M. Simonetta & S. Conway Morris (Eds.), *The Early Evolution of the Metazoa and the Significance of Problematic Taxa* (pp. 179-187). Cambridge, England: Cambridge University Press.

Howard, R. A., & Howard, E. S. (1985). "The Reverend Lansdown Guilding, 1797-1831". *Journal of the Arnold Arboretum, 37*(4), 401-402.

Hoyle, G., & del Castillo, J. (1979). Neuromuscular transmission in Peripatus. *Journal of Experimental Biology, 83*, 13-29.

Hoyle, G., & Williams, M. (1980). The musculature of Peripatus and its innervation. *Philosophical Transactions of the Royal Society B, Biological Sciences, 288*, 481-510.

Hutchinson, G. E. (1930). Restudy of some Burgess Shale fossils. *Proceedings of the United States National Museum, 78*, 1-25.

Hutchinson, G. E. (1969). Aysheaia and the general morphology of the Onychophora. *American Journal of Science, 267*, 1062-1066.

Janvier, H. (1928). Le régime de l'Opisthopatus blainvillei. *Comptes Rendus Hebdomadaires des Seances de l'Academie des Sciences, 186*, 1748-1749.

Janssen, R., & Budd, G. E. (2017). Investigation of endoderm marker-genes during gastrulation and gut-development in the velvet worm Euperipatoides kanangrensis. *Developmental Biology, 427*(1), 155-164. DOI: 10.1016/j.ydbio.2017.04.014

Jeffery, N. W., Oliveira, I. S., Gregory, T. R., Rowell, D. M., & Mayer, G. (2012). Genome size and chromosome number in velvet worms (Onychophora). *Genetica, 140*(10-12), 497-504. DOI: 10.1007/s10709-013-9698-5

Jerez-Jaimes, J. H., & Bernal-Pérez, M. C. (2009). Taxonomía de onicóforos de Santander, Colombia y termogravimetría, calorimetría de barrido diferencial y espectroscopía infrarroja de la secreción adhesiva (Onychophora: Peripatidae). *Revista de Biologia Tropical, 57*(3), 567-588.

Johow, F. (1911). Observaciones sobre los Onicóforos Chilenos. *Boletín Museo Nacional de Historia Natural [Santiago], 3*, 79-98.

Kirk, T.W. (1883). Habitat of Peripatus novo-zealandie. *New Zealand Journal of Science, 1*, 573.

Krishnan, G. (1970). Chemical nature of the cuticle and its mode of hardening in Eoperipatus weldoni. *Acta Histochemica, 37*, 1-17.

Laat, D. (2006). *Variabilidade genética e estrutura populacional de Peripatus acacioi na estação ecológica do Tripuí, MG*. Retrieved from http://www.sidalc.net/cgi-bin/wxis.exe/?IsisScript=AGB.xis&method=post&formato=2&cantidad=1&expresion=mfn=240896

Lacorte, G. A., De Sena Oliveira, I., & Da Fonseca, C. G. (2011). Phylogenetic relationships among the Epiperipatus lineages (Onychophora: Peripatidae) from the Minas Gerais State, Brazil. *Zootaxa, 2755,* 57-65.

Lankester, E. R. (1904). On the movements of the parapodia of Peripatus, millipedes and centipedes. *Quarterly Journal of Microscopical Science, 47*, 577-582.

Lavallard, R. (1965). Étude au microscope électronique de l'épithélium tégumentaire chez Peripatus acacioi Marcus et Marcus. *Comptes Rendus Hebdomadaires des Seances de l'Academie des Sciences, 260,* 965-968.

Lavallard, R., Campiglia, S., Parisi Alvares, E., & Valle, C. M. C. (1975). Contribution a la biologie de Peripatus acacioi Marcus et Marcus (Onychophore). III. *Etude descriptive de l'habitat. Vie et Milieu, 25,* 87-118.

Lawrence, R. F. (1950). Peripatus - A living museum of antiquities. *African Wildlife, 4*, 112-120.

Lawrence, R. F. (1977). Insects, arachnids and Peripatus. In A. C. Brown (Ed.), *A History of Scientific Endeavour in South Africa* (pp. 109-131). Cape Town, South Africa: Royal Society of South Africa.

Locke, M. & Huie, P. (1977). Bismuth staining of Golgi complex is a characteristic arthropod feature lacking in Peripatus. *Nature, 270*, 341-343.

Manton, S. M. (1937). Studies on the Onychophora. IV. *Proceedings of the Royal Society of London, Part B, 124*, S41.

Manton, S. M. (1972). The evolution of arthropodan locomotory mechanisms. Part 10. Locomotory habits, morphology and evolution of the hexapod classes. *Zoological Journal of the Linnean Society, 51*, 203-400.

Marcus, E. (1937). Sobre os Onychophoros. *Arquivos do Instituto Biologico [Sao Paulo], 8*, 255-266.

Mayer, G., & Whitington, P. M. (2009). Velvet worm development links myriapods with chelicerates. *Proceedings of the Royal Society B: Biological Sciences, 276*(1673), 3571-3579. DOI: 10.1098/rspb.2009.0950

Mendes, E. G., & Sawaya, P. (1958). The oxygen consumption of "Onychophora" and its relation to size, temperature and oxygen tension. *Revista Brasileira de Biologia, 18*, 129-142.

Mesibov, B. (1998). Curious, yes, but not all that rare. *Invertebrata, 11*, 6.

Milne, L. J., & Milne, M. J. (1954). The worm that didn't turn into anything (Peripatus). *Natural History [New York, USA], 63*, 182-187.

Monge-Nájera, J. (1994a). Ecological biogeography in the phylum Onychophora. *Biogeographica, 70*, 111-123.

Monge-Nájera, J. (1994b). Reproductive trends, habitat type and body characteristics in velvet worms (Onychophora). *Revista de Biología Tropical, 42*, 611-622.

Monge-Nájera, J. (1995). Phylogeny, biogeography and reproductive trends in the Onychophora. *Zoological Journal of the Linnean Society, 114*, 21-60.

Monge-Nájera, J. (1996). Jurassic-Pliocene biogeography: Testing a model with velvet worm (Onychophora) vicariance. *Revista de Biología Tropical, 44*, 159-175.

Monge-Nájera, J. (2018). City Worms (Onychophora): why do fragile invertebrates from an ancient lineage live in heavily urbanized areas? *UNED Research Journal, 10*(1), 91-94. DOI: 10.22458/urj.v10i1.2045

Monge-Nájera, J. (2019a). *"I, astonished, discovered by chance the only specimen": the first velvet worm (Onychophora).* Columna Darwin In Memoriam, Revista de Biología Tropical, Universidad de Costa Rica. Recuperado de https://revistas.ucr.ac.cr/index.php/rbt/article/view/39056

Monge-Nájera, J. (2019b). Are the eggs - kept and fed by the mother - the ancestral form of reproduction in onychophorans? Columna Darwin In Memoriam, Revista de Biología Tropical, Universidad de Costa Rica. Retrieved from https://revistas.ucr.ac.cr/index.php/rbt/issue/view/2758

Monge-Nájera, J. (2019c). *What do we really know about how velvet worms mate?.* Columna Darwin In Memoriam, Revista de Biología Tropical, Universidad de Costa Rica. Retrieved from https://revistas.ucr.ac.cr/index.php/rbt/issue/view/2758

Monge-Nájera, J. (2019d). *Why are there no onychophorans in Cuba?* Columna Darwin In Memoriam, Revista de Biología Tropical, Universidad de Costa Rica. Retrieved from https://revistas.ucr.ac.cr/index.php/rbt/issue/view/2758

Monge-Nájera, J. (2019e). *Why do some Australian onychophorans have fantastic heads?* Columna Darwin In Memoriam, Revista de Biología Tropical, Universidad de Costa Rica. Retrieved from https://revistas.ucr.ac.cr/index.php/rbt/issue/view/2758

Monge-Nájera, J. (2019f). *Why do velvet worm spermatozoa swim for years?* Columna Darwin In Memoriam, Revista de Biología Tropical, Universidad de Costa Rica. Retrieved from https://revistas.ucr.ac.cr/index.php/rbt/issue/view/2758

Monge-Nájera, J., & Alfaro, J. P. (1995). Geographic variation of habitats in Costa Rican velvet worms (Onychophora: Peripatidae). *Biogeographica, 71*, 97-108.

Monge-Nájera, J., Barquero-González, P., & Morera-Brenes, B. (2019a). Why do velvet worm spermatozoa swim for years? *Columna Darwin In Memoriam, Revista de*

*Biología Tropical*, Universidad de Costa Rica. Recuperado de
https://revistas.ucr.ac.cr/index.php/rbt/article/view/36240

Monge-Nájera, J., Barquero-González, P., & Morera-Brenes, B. (2019b). Two ways to be a velvet worm. *Columna Darwin In Memoriam, Revista de Biología Tropical*, Universidad de Costa Rica. Recuperado de
https://revistas.ucr.ac.cr/index.php/rbt/article/view/36107

Monge-Nájera, J., Barquero-González, P., & Morera-Brenes, B. (2019c). Why are there no onychophorans in Cuba? *Columna Darwin In Memoriam, Revista de Biología Tropical*, Universidad de Costa Rica. Recuperado de
https://revistas.ucr.ac.cr/index.php/rbt/article/view/36418

Monge-Nájera, J., Barquero-González, P., & Morera-Brenes, B. (2019d). Why do onychophorans and camels need the same gestation time? *Columna Darwin In Memoriam, Revista de Biología Tropical*, Universidad de Costa Rica. Recuperado de https://revistas.ucr.ac.cr/index.php/rbt/article/view/36118

Monge-Nájera, J., Barquero-González, P., & Morera-Brenes, B. (2019e). Are the eggs - kept and fed by the mother - the ancestral form of reproduction in onychophorans? *Columna Darwin In Memoriam, Revista de Biología Tropical*, Universidad de Costa Rica. Recuperado de
https://revistas.ucr.ac.cr/index.php/rbt/article/view/36398

Monge-Nájera, J., Barquero-González, P., & Morera-Brenes, B. (2019f). What do we really know about how velvet worms mate? *Columna Darwin In Memoriam, Revista de Biología Tropical*, Universidad de Costa Rica. Recuperado de
https://revistas.ucr.ac.cr/index.php/rbt/article/view/36337

Monge-Nájera, J., Barquero-González, P., & Morera-Brenes, B. (2019g). Why do some Australian onychophorans have fantastic heads? *Columna Darwin In Memoriam, Revista de Biología Tropical*, Universidad de Costa Rica. Recuperado de
https://revistas.ucr.ac.cr/index.php/rbt/article/view/36337

Monge-Nájera, J., Barrientos, Z., & Aguilar, F. (1993). Behavior of Epiperipatus biolleyi (Onychophora: Peripatidae) under laboratory conditions. *Revista de Biología Tropical, 41*, 689-696.

Monge-Nájera, J., & Hou, X. (2000). Disparity, decimation and the Cambrian "explosion": Comparison of early Cambrian and Present faunal communities with emphasis on velvet worms (Onychophora). *Revista de Biologia Tropical, 48*(2-3), 333-351.

Monge-Nájera, J., & Lourenco, W. R. (1995). Biogeographic implications of evolutionary trends in onychophorans and scorpions. *Biogeographica, 71*, 179-185.

Monge-Nájera, J., & Morera-Brenes, B. (2015). Velvet Worms (Onychophora) in Folklore and Art: Geographic Pattern, Types of Cultural Reference and Public Perception. *British Journal of Education, Society and Behavioural Science, 10*(3), 1-9.

Monge-Nájera, J., & Xianguang, H. (2002). Experimental taphonomy of velvet worms (Onychophora) and implications for the Cambrian "explosion, disparity and decimation" model. *Revista de Biologia Tropical, 50*(3-4), 1133-1138.

Montgomery, T. H. (1900). The spermatogenesis of Peripatus (Peripatopsis) balfouri up to the formation of the spermatid. *Zoologische Jahrbücher, Abteilung für Anatomie und Ontogenie der Tiere, 14*, 277-368.

Montgomery, T. H. (1912). Complete discharge of mitochondria from the spermatozoon of Peripatus. *Biological Bulletin, 22*, 309319.

Mora, M., Herrera, A. & León, P. (1996). Analisis electroforético de las secreciones adhesivas de onicoforos del genero Epiperipatus (Onychophora: Peripatidae). *Revista de Biología Tropical, 44*, 147-152.

Morera-Brenes, B., Herrera, A., Mora, M. & León, P. (1992). Estudios genomicos de Epiperipatus biolleyi (Peripatidae, Onychophora). *Revista Brasileira de Genética (Brazilian Journal of Genetics), 15*: 91.

Morera-Brenes, B., & Monge-Nájera, J. (2010). A new giant species of placented worm and the mechanism by which onychophorans weave their nets (Onychophora: Peripatidae). *Revista de Biologia Tropical, 58*(4), 1127-1142.

Morera, B., Monge-Nájera, J., & Mora, P. C. (2018). *The Conservation Status of Costa Rican Velvet Worms (Onychophora)*. DOI: 10.20944/PREPRINTS201812.0151.V1

Morera, B., Monge-Nájera, J., & Saenz, R. (1988). Parturition in onychophorans: New record and a review. *Brenesia, 29*, 15-20.

Morris, S. C. (1977). A new metazoan from the Cambrian Burgess Shale of British Columbia. *Palaeontology, 20*(3), 623-40.

Nicolas, A. (1889). Sur les rapports des muscles et des éléments épithéliaux dans le pharynx du Péripate (Peripatus capensis). *Revue Biologique du Nord de la France, 2*, 81-98.

Oliveira, I. de S., Read, V. M. S. J., & Mayer, G. (2012). A world checklist of onychophora (velvet worms), with notes on nomenclature and status of names. *ZooKeys, 211*, 1-70.

Packard, A. S. (1898). Relations of Peripatus to insects. In A. S. Packard (Ed.), *A text-book of Entomology* (pp. 9-11). New York: The MacMillan Company.

Pass, G. (1991). Antennal circulatory organs in Onychophora, Myriapoda and Hexapoda: Functional morphology and evolutionary implications. *Zoomorphology, 110*, 145-164.

Peck, S. B. (1975). A review of the New World Onychophora with the description of a new cavernicolous genus and species from Jamaica. *Psyche, 82*, 341-358.

Peters, W. (1880). Die Variation der Fusszahl bei Peripatus capensis Grube. *Sitzungsberichte der Gesellschaft Naturforschender Freunde zu Berlin [1880]*, 165-166.

Picado, C. (1911). Sur un habitat nouveau de Peripatus. Bulletin du Museum National d'Historie Naturelle, 17, 415-416.

Poinar, G. (1996). Fossil velvet worms in Baltic and Dominican amber: Onychophoran evolution and biogeography. *Science, 273*, 1370-1371.

Name and surname of all authors separated by a comma.

Poinar, Jr. G. (2000). Fossil onychophorans from Dominican and Baltic amber: Tertiapatus dominicanus n.g., n.sp. (Tertiapatidae n.fam.) and Succinipatopsis balticus n.g., n.sp. (Succinipatopsidae n.fam.) with a proposed classification of the subphylum Onychophora. *Invertebrate Biology, 119*(1), 104-109. DOI: 10.1111/j.1744-7410.2000.tb00178.x

Porter, C. E. (1905). Lecciones de historia natural dictadas a los alumnos del 4. año de la Escuela Naval (curso de 1904): Los Onicóforos. *Revista Chilena de Historia Natural ,9*, 124-128.

Porter, C. E. (1917). Bibliografía chilena razonada de Miriápodos y Onicóforos. *Revista Chilena de Historia Natural, 21*, 52-62.

Prenant, A. (1890). Note sur les éléments séminaux d'un Peripatus. *Revue Biologique du Nord de la France, 2*, 169-174.

Read, V. M. St. J. (1988). The application of scanning electron microscopy to the systematics of the neotropical Peripatidae (Onychophora). *Zoological Journal of the Linnean Society, 93*, 187-223.

Reinhard, J., & Rowell, D. M. (2005). Social behaviour in an Australian velvet worm, Euperipatoides rowelli (Onychophora: Peripatopsidae). *Journal of Zoology, 267*(1), 1-7.

Rhebergen, F., & Donovan, S. K. (1994). A lower Palaeozoic "onychophoran" reinterpreted as a pelmatozoan (stalked echnioderm) column. *Atlantic Geology, 30*, 19-23.

Rolfe, W. D. I., Schram, F. R., Pacaud, G., Sotty, D., & Secretan, S. (1982). A remarkable Stephanian biota from Montceau-les-Mines, France. *Journal of Paleontology, 56*, 426-428.

Ruhberg, H. (1979). Light and electron microscopical investigations on the salivary glands of Opisthopatus cinctipes and Peripatopsis moseleyi (Onychophora: Peripatopsidae). *Zoologischer Anzeiger, 203*, 35-47.

Ruhberg, H. (1985). Die Peripatopsidae (Onychophora). Systematik, Ökologie, Chorologie und phylogenetische Aspekte. In F. Schaller (Ed.), *Zoologica*. Stuttgart: Schweizerbart'sche Verlagsbuchhandlung.

Ruhberg, H., Tait, N. N., Briscoe, D. A. & Storch, V. (1988). Cephalofovea tomahmontis n. gen., n. sp., an Australian peripatopsid (Onychophora) with a unique cephalic pit. *Zoologischer Anzeiger, 221*, 117-133.

Ryu, S. H., Kang, C. W., Choi, J., Myung, Y., Ko, Y.J., Lee, S. M., … Son, S. U. (2018). Microporous Porphyrin Networks Mimicking a Velvet Worm Surface and Their Enhanced Sensitivities toward Hydrogen Chloride and Ammonia. *ACS Applied Materials & Interfaces, 10*(8), 6815-6819. DOI: 10.1021/acsami.7b19119

Saint-Remy, G. (1889). Sur la structure du cerveau chez les Myriopodes et les Arachnides. *Revue Biologique du Nord de la France, 2*, 41-55.

Sanchez, S. (1958). Cellules neurosécrétrices et organes infracérébraux de Peripatopsis moseleyi Wood (Onychophores) et neurosécrétion chez Nymphon gracile Leach (Pycnogonides). *Archives de Zoologie Experimentale et Generale [Notes et Revue], 96*, 57-62.

Santana, G. G., Almeida, W. D. O., Alves, R. R. da N., & Vasconcellos, A. (2008). Extension of the northern distribution of Onychophora in the Brazilian Atlantic Forest. *Biotemas, 21*(2), 161-163. DOI: 10.5007/2175-7925.2008v21n2p161

Schürmann, F. W., & Sandeman, D. C. (1976). Giant fibres in the ventral nerve cord of Peripatoides leuckarti (Onychophora). *Naturwissenschaften, 63*, 580-581.

Sclater, W. L. (1888). On the early stages of the development of South American species of Peripatus. *Quarterly Journal of Microscopical Science, 28*, 343-363.

Sedgwick, A. (1884). On the origin of metameric segmentation and some other morphological questions. *Quarterly Journal of Microscopical Science, 24*, 43-82.

Sedgwick, A. (1908a). The distribution and classification of the Onychophora. *Quarterly Journal of Microscopical Science, 52*, 379-406.

Sedgwick, A. (1908b). Relation between the geographical distribution and the classification of the Onychophora. *Proceedings of the Cambridge Philosophical Society, 14*, 546.

Sheldon, L. (1887). On the development of Peripatus novae-zealandiae. *Quarterly Journal of Microscopical Science, 28*, 205-237.

Sheldon, L. (1890). The maturation of the ovum in the Cape and New Zealand species of Peripatus. *Quarterly Journal of Microscopical Science, 30*, 1-29.

Silva, J. R. M. C. da, Coelho, M. P. D., & Nogueira, M. I. (2000). Processo inflamatório induzido em peripatus acacioi (onychophora) um fóssil vivo. *Journal Invertebrate Pathology, 75*, 41-46.

Simoes, L. C. G., Marques da Silva, I., & Schreiber, G. (1964). DNA e volume nuclear em tecidos de Peripatus acacioi (Marcus & Marcus). *Ciencia e Cultura [Sao Paulo], 16*, 291-295.

Snodgrass, R. E. (1938). Evolution of the Annelida, Onychophora and Arthropoda. *Smithsonian Miscellaneous Collections, 97*, 1-159.

Sosa-Bartuano, Á., Monge-Nájera, J., & Morera-Brenes, B. (2018). A proposed solution to the species problem in velvet worm conservation (Onychophora). *UNED Research Journal, 10*(1), 204-208.

Storch, V., & Ruhberg, H. (1977). Fine structure of the sensilla of Peripatopsis moseleyi (Onychophora). *Cell & Tissue Research, 177*, 539-553.

Storch, V., Alberti, G. & Lavallard, R., & Campiglia, S. S. (1979). Données ultrastructurales sur des zones synaptiques dans les nephridies de Peripatus acacioi Marcus et Marcus (Onychophore). *Boletim de Fisiologia Animal, 3*, 145-156.

Tait, N. N., & Norman, J. M. (2001). Novel mating behaviour in Florelliceps stutchburyae gen. nov., sp. nov. (Onychophora: Peripatopsidae) from Australia. *Journal of Zoology, 253*(3), 301-30.

Tarlo, L. B. (1967). Xenusion - Onychophoran or Coelenterate? *Mercian Geologist, 2*, 97-99.

Thompson, I., & Jones, D. S. (1980). A possible onychophoran from the Middle Pennsylvanian Mazon Creek Beds of northern Illinois. *Journal of Paleontology, 54*, 588-596.

Treffkorn, S., & Mayer, G. (2017). Conserved versus derived patterns of controlled cell death during the embryonic development of two species of Onychophora (velvet worms). *Developmental Dynamics, 246*(5), 403-416. DOI: 10.1002/dvdy.24492

Trindade, G. (1958). Peripatus e sua possivel utilizacao em laboratorio. *Anais do III Congresso de Farmacia e Bioquimica Pan-americano e V Congresso Brasileiro de Farmacia* 519-520.

Tuzet, O., & Manier, J. F. (1958). Recherches sur Peripatopsis moseleyi Wood-Mason péripate du Natal. I. Étude sur le sang. II. La spermatogenèse. *Bulletin Biologique de la France et de la Belgique, 92*, 7-23.

Vachon, M. (1954). Répartition actuelle et ancienne des Onychophores ou Peripates. *Revue Générale des Sciences Pures et Appliquées et Bulletin de l'Association Francaise pour l'Avancement des Sciences, 61*, 300-308.

van der Lande, V.M. (1991). Native and introduced Onychophora in Singapore. *Zoological Journal of the Linnean Society, 102*, 101-114.

Vincent, M. (1936). Un nouveau type de Coccidie des Péripates. *Comptes Rendus des Seances de la Société de Biologie et de ses Filiales, 122*, 260-262.

Wägele, J. W. (1993). Rejection of the "Uniramia" hypothesis and implications of the Mandibulata concept. *Zoologische Jahrbücher, Abteilung für Systematik, Ökologie und Geographie der Tiere, 120*, 253-288.

Walcott, C. D. (1931). Addenda to descriptions of Burgess shale fossils. *Smithsonian Miscellaneous Collections, 85*, 1-46.

Wells, S. M., Pyle, R. M., & Collins, N. M. (1983). Peripatus. Phylum Onychophora. In *The IUCN Invertebrate Red Data Book IUCN* (pp. 515-520). Gland, Switzerland.

Wenzel, R. L. (1950). Peripatus - "living fossil" and "missing link". *Tuatara, 3*, 98-99.

Whitman, C.O. (1891). Spermatophores as a means of hypodermic impregnation. *Journal of Morphology, 4*, 361-406.

Wood-Mason, J. (1879). Morphological notes bearing on the origin of insects. *Transactions of the Entomological Society of London [1879], 2*, 145-167.

Zilch, A. (1955). Begegnung mit Peripatus. *Natur und Volk, 85*, 146-154.

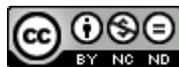